\documentclass[twocolumn,showpacs,preprintnumbers,amsmath,amssymb]{revtex4}

\usepackage{graphicx}
\usepackage{dcolumn}
\usepackage{bm}

\begin{document}


\title{Frustrated quantum-spin system on a triangle coupled with $e_g$
lattice vibrations - Correspondence to Longuet-Higgins {\it et al.}'s Jahn-Teller model -}
\author{Hisatsugu  Yamasaki}
\email{hisa@physics.s.chiba-u.ac.jp}
\homepage{http://zeong.s.chiba-u.ac.jp/~hisa/}
\author{Yuhei Natsume}%
\affiliation{%
Graduate School of Science and Technology, Chiba-University\\
Inage-ku, Chiba, 263-8522 Japan
}%

\author{Akira Terai}
\author{Katsuhiro Nakamura}
\affiliation{
Department of Applied Physics,
Osaka City University\\
Sumiyoshi-ku
Osaka, 558-8585 Japan
}%
     
\date{\today}

\begin{abstract}
We investigate the quantum three spin model $({\bf S_1},{\bf S_2},{\bf S_3})$
 of spin$=1/2$ on a triangle, in which spins are coupled with lattice-vibrational
 modes through the exchange interaction depending on distances between
 spin sites. The present model corresponds to the dynamic Jahn-Teller system
 $E_g\otimes e_g$ proposed by Longuet-Higgins {\it et al.}, Proc.R.Soc.A.{\bf
 244},1(1958). This correspondence is revealed by using the transformation to
 Nakamura-Bishop's bases proposed in Phys.Rev.Lett.{\bf
 54},861(1985). Furthermore, we elucidate the
 relationship between the behavior of a chiral order parameter ${\hat \chi}={\bf S_1\cdot(S_2\times S_3)}$ and that of the electronic orbital angular momentum ${\hat \ell_z}$ in $E_g\otimes e_g$ vibronic model: The regular oscillatory behavior of the expectation value $\langle {\hat \ell_z} \rangle$ for vibronic structures with increasing energy can also be found in that of $\langle {\hat \chi} \rangle$. The increase of the additional anharmonicity(chaoticity) is found to yield a rapidly decaying irregular oscillation of $\langle {\hat \chi} \rangle$.
     
\end{abstract}

\maketitle

\newcommand{\obs}[1]{\ensuremath{\overrightarrow{\boldsymbol{#1}}}}
\newcommand{\bol}[1]{\ensuremath{\boldsymbol{#1}}}
\newcommand{\suu}{\ensuremath{|\uparrow\uparrow\rangle}}
\newcommand{\sud}{\ensuremath{|\uparrow\downarrow\rangle}}
\newcommand{\sdu}{\ensuremath{|\downarrow\uparrow\rangle}}
\newcommand{\sdd}{\ensuremath{|\downarrow\downarrow\rangle}}
\newcommand{\suuu}{\ensuremath{|\uparrow\uparrow\uparrow\rangle}}
\newcommand{\suud}{\ensuremath{|\uparrow\uparrow\downarrow\rangle}}
\newcommand{\sudu}{\ensuremath{|\uparrow\downarrow\uparrow\rangle}}
\newcommand{\sudd}{\ensuremath{|\uparrow\downarrow\downarrow\rangle}}
\newcommand{\sduu}{\ensuremath{|\downarrow\uparrow\uparrow\rangle}}
\newcommand{\sdud}{\ensuremath{|\downarrow\uparrow\downarrow\rangle}}
\newcommand{\sddu}{\ensuremath{|\downarrow\downarrow\uparrow\rangle}}
\newcommand{\sddd}{\ensuremath{|\downarrow\downarrow\downarrow\rangle}}
\newcommand{\bsuuu}{\ensuremath{\langle\uparrow\uparrow\uparrow|}}
\newcommand{\bsuud}{\ensuremath{\langle\uparrow\uparrow\downarrow|}}
\newcommand{\bsudu}{\ensuremath{\langle\uparrow\downarrow\uparrow|}}
\newcommand{\bsudd}{\ensuremath{\langle\uparrow\downarrow\downarrow|}}
\newcommand{\bsduu}{\ensuremath{\langle\downarrow\uparrow\uparrow|}}
\newcommand{\bsdud}{\ensuremath{\langle\downarrow\uparrow\downarrow|}}
\newcommand{\bsddu}{\ensuremath{\langle\downarrow\downarrow\uparrow|}}
\newcommand{\bsddd}{\ensuremath{\langle\downarrow\downarrow\downarrow|}}
The triangular Heisenberg antiferromagnets play an important role in our
understanding the resonating valence bond(RVB) state, in which the scalar 
chirality for three spins
${\bf S_1\cdot(S_2\times S_3)}$ is expected to have a nonzero expectation
value\cite{1,2,3,14}.
This subject has been a focus of recent experimental activities\cite{11,12,5}, since it
was expected that a frustrated $s=1/2$ triangular antiferromagnet lattice
might be realized in NaTiO${}_2$ and LiNiO${}_2$\cite{4}. 

In this Letter, we propose a triangular cluster model of
the Heisenberg antiferromagnet in which quantum spins are
coupled with lattice vibrations, for the purpose to discuss magnetic properties in
relation to the typical dynamical Jahn-Teller system $E_g\otimes e_g$. In short, 
the spin-lattice interaction is introduced by expanding the exchange
interaction with respect to deviation of lattice displacements from equilibrium. We shall address to the following issue: With use of a
unitary transformation for this spin system, the proposed model becomes equivalent to that of the well-known vibronic problem for $E_g\otimes
e_g$ Jahn-Teller system\cite{7}.

\begin{figure}[h]
  \begin{minipage}{.25\textwidth}
   \includegraphics[width=\linewidth]{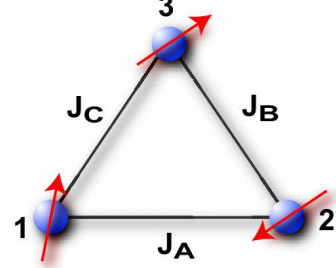}
  \end{minipage}
  \caption{Triangle with antiferromagnetic spins.}
\label{fig1}
\end{figure}

Let us consider the quantum spin system where three spins of spin=$1/2$ are
localized at lattice sites 1,2 and 3 on triangle. The coupling between neighboring spins are expressed by the antiferromagnetic exchange
interactions $J_A$, $J_B$ and $J_C$ as shown in Fig.\ref{fig1}. 
The
corresponding Heisenberg Hamiltonian is 
\begin{equation}
 \mathcal{H}=J_A{\bf S_1\cdot S_2} + J_B{\bf S_2\cdot S_3} + J_C{\bf S_3\cdot S_1}.
\end{equation}
We concentrate our attention on the spin state where $z$ component of the total spin satisfies
$s_{1z}+s_{2z}+s_{3z}=1/2$. Therefore, these bases are expressed explicitly as $\sduu,\sudu,\suud$,
where arrows denote $s_{jz}$ for site $j$. 
By using these bases, we obtain the exchange Hamiltonian,
\begin{widetext}
\begin{equation}
 \mathcal{H}/\left(-\frac{\hbar^2}{4}\right)=  
\bordermatrix{
        & \sduu     &    \sudu      &   \suud      \cr
 \bsduu  &-J_A+J_B-J_C &    2J_A       &    2J_C      \cr
 \bsudu  &  2J_A     & -J_A-J_B+J_C  &    2J_B      \cr
 \bsuud  &  2J_C     &    2J_B       & J_A-J_B-J_C  \cr
}
.\label{spin}
\end{equation}
\end{widetext}

\begin{figure}[htbp]
\begin{minipage}{.20\textwidth}
   \includegraphics[width=\linewidth]{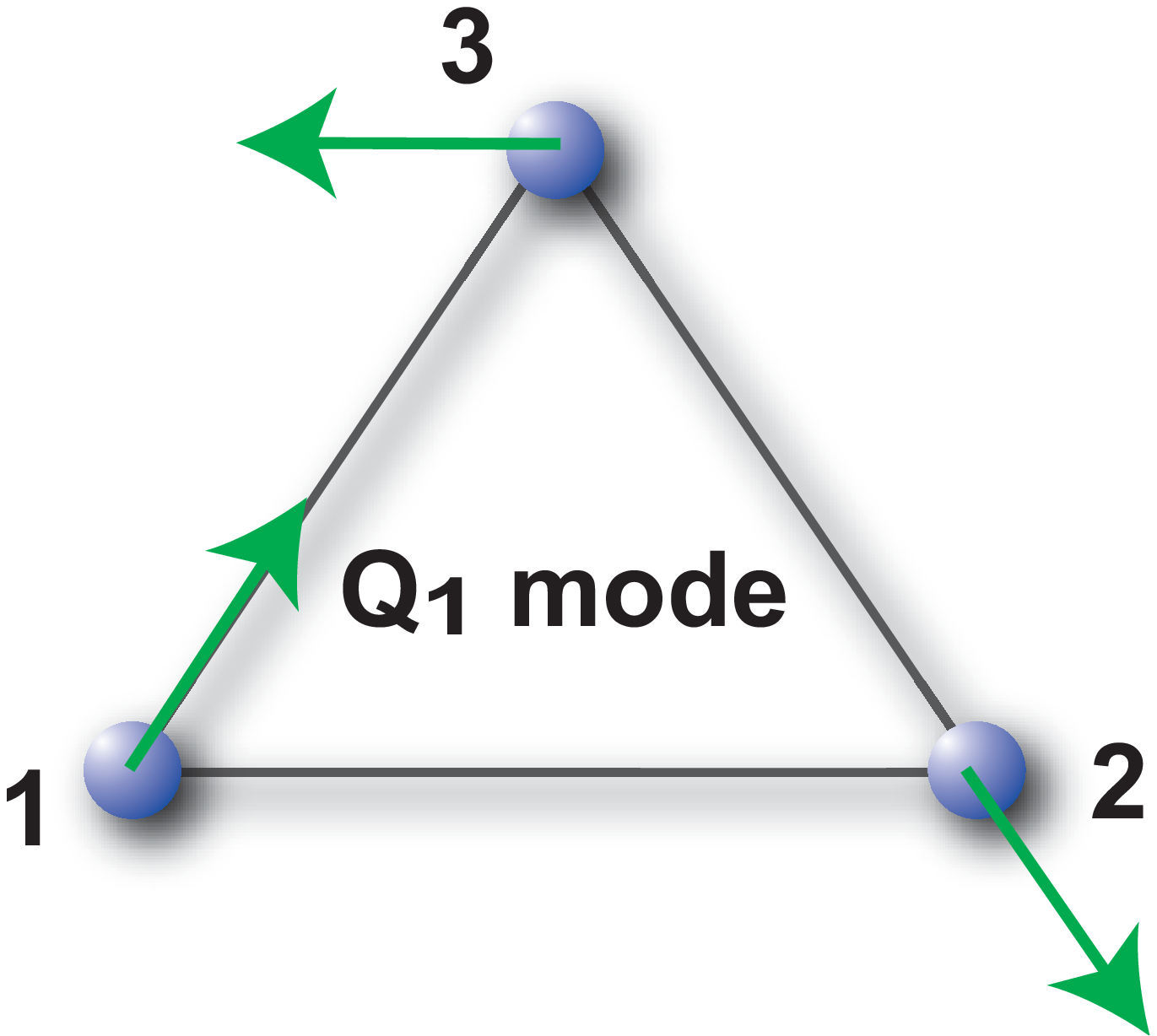}
  \end{minipage}
\hfill
  \begin{minipage}{.20\textwidth}
   \includegraphics[width=\linewidth]{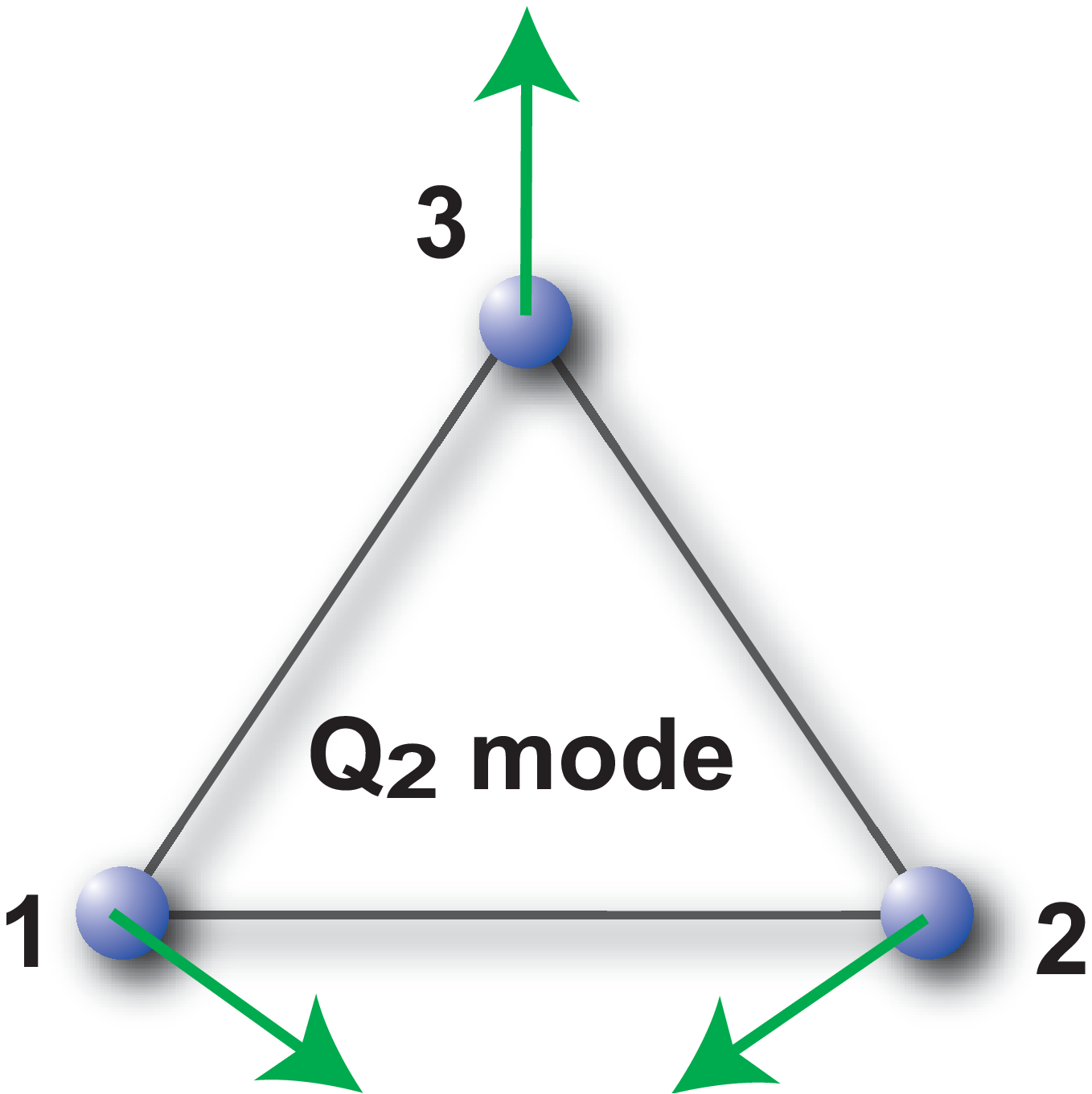}
  \end{minipage}
\caption{The normal modes $Q_1$ and $Q_2$ in the triangle.}
\label{fig2}
\end{figure}

Next we introduce the interaction
between the spins and lattice vibrations, noting the dependence of $J_A,J_B$ and
$J_C$ on the distances between spin sites. As for the lattice vibration, we
employ the normal modes for the triangle; The normal $e_g$ modes $Q_1$ and
$Q_2$, which are degenerate, are given in
Fig.\ref{fig2}. 
The remaining $a_{1g}$ mode(:the breathing mode) has a much higher strain
energy and is ignored hereafter. (There are other global degrees of freedom
related to translation of the center of mass and to rotation around the axis
perpendicular to the triangular plane. They however have nothing to do with
lattice vibrations and are also ignored.) Then the spin-lattice interaction 
is obtained as a result of
the the expansion of $J_A,J_B$ and $J_C$ linearly in the $e_g$ modes as follows:
\begin{eqnarray}
 J_A &=& J\cdot\left[1+\frac{\alpha}{2}(Q_1-\sqrt{3}Q_2)\right] \nonumber \\
 J_B &=& J\cdot\left[1-\alpha Q_1\right] \label{couple} \\
 J_C &=& J\cdot\left[1+\frac{\alpha}{2}(Q_1+\sqrt{3}Q_2)\right], \nonumber
\end{eqnarray}
where $\alpha$ is the coupling constant. 

Concerning the spin system, on the other hand, we introduce the following
bases introduced by Nakamura and Bishop for the triangular spin plaquet\cite{33,34,35}:
\begin{widetext}
\begin{eqnarray}
 \arrowvert k=0\rangle               &=& \frac{1}{\sqrt{3}}\left(\sduu + \sudu + \suud \right) \nonumber \\
 \arrowvert k=\frac{2\pi}{3}\rangle  &=& \frac{1}{\sqrt{3}}\left(\sduu +
													   e^{\frac{2\pi}{3} i}\sudu + e^{-\frac{2\pi}{3} i}\suud\right) \label{base}\\
 \arrowvert k=-\frac{2\pi}{3}\rangle &=& \frac{1}{\sqrt{3}}\left(\sduu +
													   e^{-\frac{2\pi}{3} i}\sudu + e^{\frac{2\pi}{3}i}\suud\right).\nonumber
\end{eqnarray}
\end{widetext}
These bases reflect clockwise and anticlockwise rotations of a spin
configuration on the plane of the
triangle. The wave numbers $k=0,\pm 2\pi/3$ correspond to phase factors in
Bloch's theorem for the system with the discrete rotational symmetry.
From the viewpoint of the ligand-field theory\cite{6}, the construction of these bases (\ref{base}) from $\sduu,\sudu$ and $\suud$ is regarded as a
formation of $E_g$ and $A$ representations in $D_{3d}$ symmetry from the
triply-degenerate $T_{2g}$ ones in $O_h$ symmetry.
By using this new bases, the Hamiltonian matrix (\ref{spin}) can be transformed to 
\begin{widetext}
\begin{equation}
 \mathcal{H}/\left(-\frac{3}{4}\hbar^2J\right)=
\bordermatrix{
       & |k=0\rangle & |k=\frac{2\pi}{3}\rangle & |k=-\frac{2\pi}{3}\rangle \cr 
     \langle k=0|   & 1 &                            0       &    0      \cr
     \langle k=\frac{2\pi}{3}|  & 0 &                         -1         & \alpha(-Q_1-iQ_2)         \cr
     \langle k=-\frac{2\pi}{3}| & 0 &   \alpha(-Q_1+iQ_2)  &  -1
}.\label{trans}
\end{equation}
\end{widetext}
From Eq.(\ref{trans}) we find that the $k=0$ manifold is completely separated from other manifolds, i.e., $\mathcal{H}=\mathcal{H}_{k=0} \otimes \mathcal{H}_{k=\pm 2\pi/3}$. $\mathcal{H}_{k=0}$ and $\mathcal{H}_{\pm2\pi/3}$ correspond to $A$ and $E_g$ representation, respectively.
The interaction Hamiltonian $\mathcal{H}_{k=\pm2\pi/3}$ can result in a pair of adiabatic energy surfaces, which together with the harmonic term $(\propto Q_1^2+Q_2^2)$, forms the Mexican hat potential. In fact, by applying the unitary transformation:
\begin{equation}
 U=-\frac{1}{\sqrt{2}}
\left( \begin{array}{cc}
       1 &       i            \\
      -1 &       i         
\end{array} \right),\label{unitary}
\end{equation}
we can get
\begin{eqnarray}
 \mathcal{{\tilde H}}_{k=\pm2\pi/3} &=& U^{-1}\mathcal{H}_{k=\pm2\pi/3}U \nonumber \\
 &=& \frac{3}{4}\hbar^2J{\bf I}-\frac{3\alpha}{4}\hbar^2J
\left( \begin{array}{cc}
       Q_1   &       +Q_2            \\
       +Q_2 &       -Q_1         
\end{array} \right)
.\label{mat}
\end{eqnarray}
This expression accords with the electron-lattice interaction part of the vibronic
Hamiltonian for the Jahn-Teller system $E_g \otimes e_g$, 
\begin{equation}
 \mathcal{H}_{JT} = \frac{1}{2}\omega^2(Q_1^2+Q_2^2)+\alpha^{'}
\left( \begin{array}{cc}
       Q_1   &       +Q_2            \\
       +Q_2 &       -Q_1         
\end{array} \right).\label{JT}
\end{equation}
Thus, we would like to emphasize that the present system for quantum spins on
the triangle coupled with doubly-degenerate vibrational $e_g$ modes is equivalent
to the vibronic model for the $E_g\otimes e_g$ system that is extensively and intensively investigated in the context of the dynamical Jahn-Teller problem.

Before proceeding to the argument on the chiral order parameter of the spin system, we shall recall the definition of the electronic orbital angular momentum in the dynamical Jahn-Teller system.
For the Hamiltonian consisting of the
kinetic energy ($1/2(P_1^2+P_2^2)$) and $\mathcal{H}_{JT}$ of (\ref{JT}), the
$p$-th eigenstate of the $\ell=1/2$ manifold, $\Psi_{p,1/2}$ is given by 
\begin{widetext}
\begin{equation}
 \Psi_{p,1/2}=a_{1,p}\psi_{1,0}\phi_{+}+a_{2,p}\psi_{2,1}\phi_{-}+a_{3,p}\psi_{3,0}\phi_{+}+a_{4,p}\psi_{4,1}\phi_{-}+\ldots \label{wave}
\end{equation}
\end{widetext}
where the $\psi_{n,m}$'s are the eigenfunctions of the
isotropic two-dimensional harmonic oscillator ($n$ and $m$ are radial and
azimuthal quantum numbers, respectively), and $\phi_+$ and $\phi_-$
are degenerate electronic states $\phi_{\pm}=d_{u}\pm id_{v}$.
The expansion (\ref{wave}) was found by rewriting $\mathcal{H}_{JT}$ in
(\ref{JT}) into a suitable form with the use of $\phi_{\pm}$\cite{32}.
In the context of the spin-lattice system under consideration, the block
matrix $\mathcal{H}_{k=\pm 2\pi/3}$ in (\ref{trans}) already takes such a
suitable form with the use of Nakamura-Bishop's bases $|k=\pm 2\pi/3\rangle$, and the vibronic
wave function takes the same form as (\ref{wave}).

In the vibronic state $\Psi_{p,1/2}$ in the dynamical Jahn-Teller system, the
expectation value of the electronic orbital angular momentum
${\hat \ell_{z}}$ is given as 
\begin{eqnarray}
 \langle {\hat \ell_{z}} \rangle_p &=& \langle \Psi_{p,1/2} | {\hat \ell_{z}} | \Psi_{p,1/2} \rangle \nonumber \\   
&=& \sum_{n=1}^{\infty}|a_{n,p}|^2(-1)^{n-1}\Xi_{\ell} \qquad (p=1,2,\ldots). \label{func}
\end{eqnarray}
Here, $\Xi_{\ell}$ is the expectation value of ${\hat \ell_z}$
in the electronic states $\phi_{+}$ and $\phi_{-}$: 
\begin{equation}
 \Xi_{\ell}=\langle \phi_{+} | {\hat \ell_{z}} | \phi_{+} \rangle = - \langle \phi_{-} | {\hat \ell_{z}} | \phi_{-} \rangle . \label{elect}
\end{equation}
The emergence of an outstanding regular oscillation of $\langle {\hat \ell_{z}}
\rangle_p$ as a function of energy$(p)$ was pointed out three decades
ago\cite{17}, which has received a renewed attention recently in the context
of nonlinear dynamics\cite{32}.

\begin{figure*}[htbp]
  \begin{minipage}{.47\textwidth}
   \includegraphics[width=\linewidth]{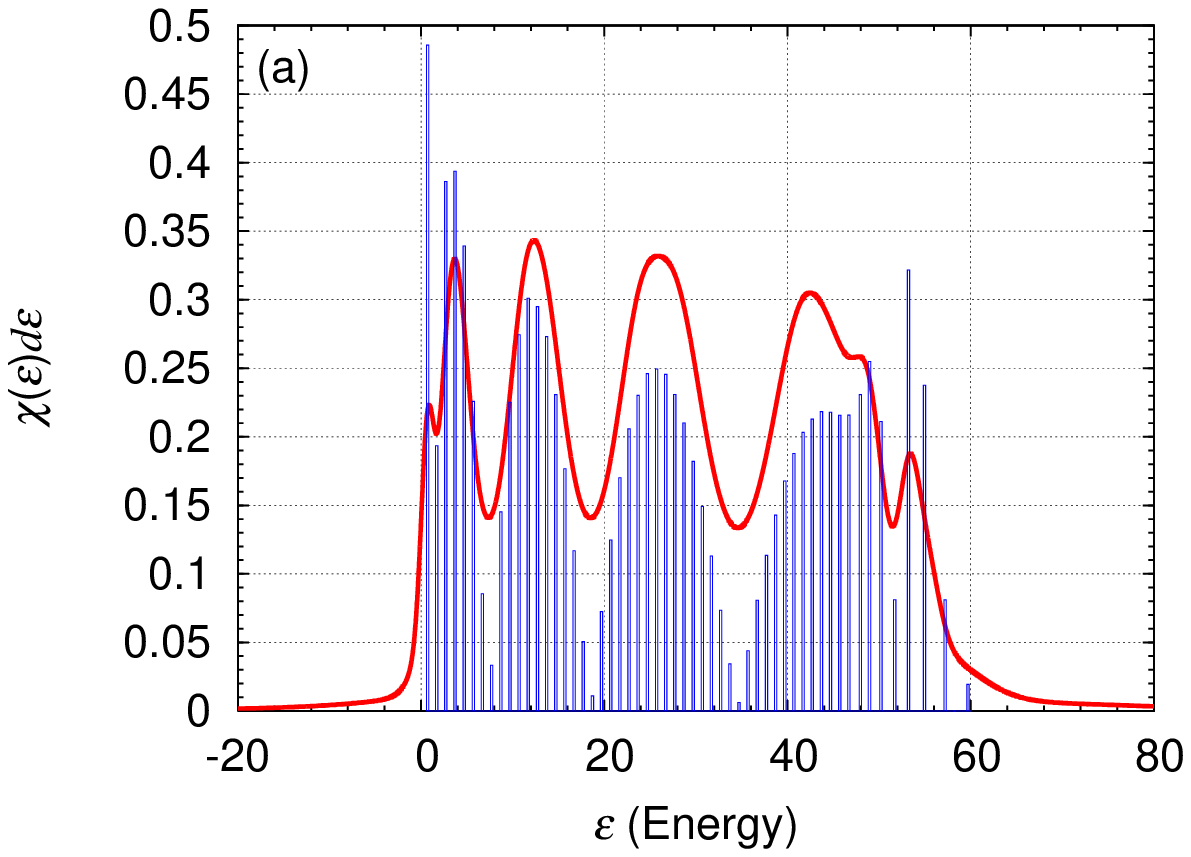}
  \end{minipage}
\hfill
  \begin{minipage}{.47\textwidth}
   \includegraphics[width=\linewidth]{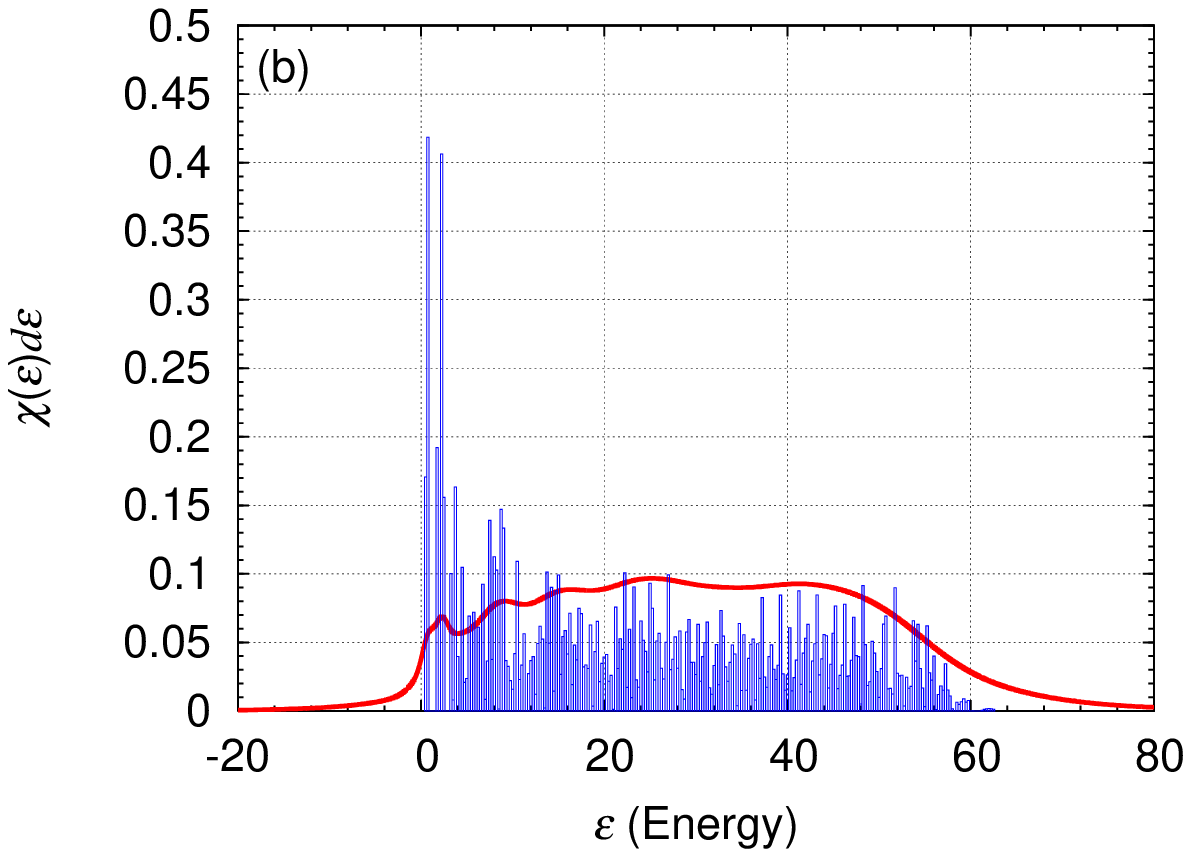}
  \end{minipage}
  \caption{Energy($\varepsilon$) dependence of partially-averaged chirality
  $\chi(\varepsilon)d\varepsilon (=\sum_p^{'} | \sum_{n=1}^{\infty}(-1)^na_{n,p}^2 | d\varepsilon)$ with $\varepsilon=0.25$ in unit of $\Xi_{\chi}$. (a) and (b) correspond to $\gamma=0,1.41$, respectively. $\alpha=0.707$ and the unit of energy is $\hbar\omega$. Envelop functions are also constructed by Gaussian coarse-graining of each peak. (Reproduced from \cite{32} and interpreted in the context of a triangular spin model.)}
\label{fig3}
\end{figure*}

Now let's come back to the argument of the characteristic operator for the quantum spin system. With use of the bases (\ref{base}), we evaluate the expectation values for
chiral order parameter 
\begin{equation}
 {\hat \chi}={\bf S_1 \cdot (S_2\times S_3)}.\label{chi}
\end{equation}
The order parameter ${\hat \chi}$ characterizes degree of the frustration of the triangular antiferromagnet\cite{14}.
The expectation values of ${\hat \chi}$ in each of the $E_g$ states (\ref{base}) are 
\begin{eqnarray}
 \langle k=\frac{2\pi}{3} | {\hat \chi} | k=\frac{2\pi}{3} \rangle &=& -\frac{\sqrt{3}}{4}\equiv -\Xi_{\chi} \nonumber \\
 \langle k=-\frac{2\pi}{3} | {\hat \chi} | k=-\frac{2\pi}{3} \rangle &=& \frac{\sqrt{3}}{4}\equiv \Xi_{\chi}. \label{expect}
\end{eqnarray}
(The value $ \langle k=0 | {\hat \chi} | k=0 \rangle = 0$ is now irrelevant
since $|k=0\rangle$ is coupled only with the higher frequency $a_{1g}$ mode.)
Thus, the states  $ | k=\pm\frac{2\pi}{3} \rangle$ and chiral order parameter
${\hat \chi}$ in the spin-lattice system correspond to states
$|\phi_{\pm} \rangle$ and ${\hat \ell_z}$ in the dynamical Jahn-Teller
system, respectively. Taking the eigenstates similar to (\ref{wave}), the dependence
of $\langle {\hat \chi} \rangle_p$ on the $p$-th eigenstate is given by 
\begin{equation}
 \langle {\hat \chi} \rangle_p = \sum_{n=1}^{\infty} |a_{n,p}|^2(-1)^n \Xi_{\chi}. \label{chi2}
\end{equation}
This means that the behavior of $\langle {\hat \chi} \rangle_{p}$ can be
revealed by applying the analysis of $\langle {\hat \ell_z} \rangle_{p}$
in (\ref{func}). In fact, the expectation values $\langle
{\hat \chi} \rangle_p$ in (\ref{chi2}) shows regular oscillation with increasing the energy(see Fig.\ref{fig3}(a)), just as in the case of $\langle
{\hat \ell_z} \rangle_p$ in the dynamical Jahn-Teller system\cite{17}.

Finally we note a role of the anharmonic term involved in the triangular
three particle system. Let us introduce Toda-lattice potential\cite{10} 
\begin{equation}
 U(x)=\frac{c}{d}e^{-d x}+c x-\frac{c}{d},
\end{equation}
where $x$ is the deviation of inter-particle distance from the equilibrium lattice constant. $c$ and $d$ are
constant with the condition of $cd>0$. The total lattice potential is a sum of
$U(x)$ with $x$ the three kind of deviations for three segments of the regular
triangle. In the limit $d<<1$ under the constraint $cd=$ constant, we obtain
the following expansion in $x$: 
\begin{eqnarray}
 U(x)&=&\frac{c}{d}(1-d x+\frac{d^2}{2!}x^2-\frac{d^3}{3!}x^3+\ldots)+c x-\frac{c}{d} \nonumber \\
&=& \frac{c d}{2}x^2-\frac{c d^2}{6}x^3+\ldots . \label{pot}
\end{eqnarray}
Suppressing a high-frequency $a_{1g}$ mode and noting the symmetry of the $e_g$ modes in Fig.{\ref{fig2}}, the bilinear term
in (\ref{pot}) leads to the 2-d
harmonic oscillator potential. On the other hand, the cubic term in (\ref{pot}) leads to the
trigonal(anharmonic) potential
\begin{equation}
 V_A=V_A(Q_1,Q_2)=-\frac{\gamma}{3}(Q_1^3-3Q_1Q_2^2)
\end{equation} 
with $\gamma=cd^2/2$ in terms of normal $e_g$ modes $Q_1$ and $Q_2$. The
chaotic semiclassical dynamics induced by the anharmonic
term was explored in the context of the dynamical Jahn-Teller
system\cite{32}, 
and we can expect a rapidly decaying irregular
oscillation of $\langle {\hat \chi} \rangle$ by increasing the
anharmonicity(chaoticity)(see Fig.\ref{fig3}(b)).

In conclusion the frustrated quantum spin system on a triangle coupled with
lattice vibrations is equivalent to $E_g\otimes e_g$ Jahn-Teller system. The chiral order parameter ${\hat
\chi}$ should signify a
quantum chaos induced by the coupling between quantum
spins and lattice vibrations, and the energy dependence of $\langle {\hat \chi}
\rangle$ shows the transition from regular to irregular oscillations by
adding the anharmonicity.


\begin{thebibliography}{9}
\bibitem{1} G.Baskaran, Phys.Rev.Lett.{\bf 63}(1998)2524.
\bibitem{2} P.W.Anderson, Material Res. Bull.{\bf 8}(1973)153.
\bibitem{3} P.Fazekas and P.W.Anderson, Phil.Mag.{\bf 30}(1974)432.
\bibitem{14} H.Kawamura, J.Phys.:Condens.Matter{\bf 10}(1998)4707.
\bibitem{11} A.Lascialfari {\it et al.}, Phys.Rev.B{\bf 67}(2003)224408.
\bibitem{12} V.P.Plakhty {\it et al.}, Phys.Rev.Lett.{\bf 85}(2000)3942.
\bibitem{5} Y.Kitaoka {\it et al.}, J.Phys.Soc.Jpn.{\bf 67}(1998)3703.
\bibitem{4} K.Hirakawa,  H.Kadowaki  and K.Ubukoshi,\\ J.Phys.Soc.Jpn.{\bf
		54}(1985)3526; K.Takeda {\it et al.},\\ J.Phys.Soc.Jpn.{\bf 61}(1992)2156.
\bibitem{7} H.C.Longuet-Higgins, U.{\"O}pik and M.H.L.Pryce,\\ Proc.\ Roy.\ Soc.\ A.{\bf
		244},(1958)1.
\bibitem{33} K.Nakamura and A.R.Bishop,Phys.Rev.Lett.{\bf 54}(1985)861.
\bibitem{34} K.Nakamura and A.R.Bishop,Phys.Rev.B.{\bf 33}(1986)1963.
\bibitem{35} K.Nakamura, Quantum Chaos{\it -A New Paradigm of Nonlinear
		Dynamics-}(Cambridge University Press, Cambridge, 1993).
\bibitem{6} S.Sugano and Y.Tanabe and H.Kamimura, {\it Multiplets of
		Transition-Metal Ions in Crystals}(Academic Press,1970).
\bibitem{32} H. Yamasaki, Y. Natsume, A. Terai and K. Nakamura,
		Phys.Rev.E. {\bf 68}(2003)046201.
\bibitem{17} S.Washimiya, Phys.Rev.Lett.{\bf 28}(1972)556.
\bibitem{10} M.Toda, J.Phys.Soc.Jpn.{\bf 22}(1967)431;{\bf 23}(1967)501.
\end{thebibliography}
\end{document}